\documentclass[12pt]{article}
\usepackage{slashed}
\usepackage{amsmath}
\usepackage{graphicx}
\begin{document}

 \title{\bf Entropy in quantum chromodynamics}
\author{John M. Cornwall, Dept. of Physics and Astronomy, UCLA}
\date{February 6, 2012}
\maketitle

\begin{abstract}
We review the role of zero-temperature entropy in several closely-related contexts in   QCD.  The first is  entropy associated with disordered condensates, including $\langle G_{\mu\nu}^2\rangle$.  The second is  effective vacuum entropy arising from QCD solitons such as center vortices,  yielding  confinement and chiral symmetry breaking.  The third is   entanglement entropy, which is entropy associated with a pure state, such as the QCD vacuum, when the state is partially unobserved and unknown.  Typically,  entanglement entropy of an unobserved 3-volume   scales not with the volume but with the area of its bounding surface.   The fourth manifestation of entropy in QCD is the   configurational entropy of light-particle world lines and flux tubes; we argue that this entropy is critical for understanding how confinement produces  chiral symmetry breakdown, as manifested by a dynamically-massive quark, a massless pion, and a $\langle \bar{q}q\rangle$ condensate.

\end{abstract}

\section{Introduction}

Entropy is a precise tool for specifying imprecision.  In statistical mechanics and in certain areas of  field theory, the entropy $S(q_i)$, a function of a small number of state variables $q_i$, is defined as the logarithm of the number $\mathcal{N}\gg 1$ of system states consistent with the small number of state variables:  $S=\ln \mathcal{N}$.  The imprecision arises because all the system states have essentially the same values of the state variables, such as (in thermodynamics) total internal energy $U$ and particle number $N$, and one does not know or care precisely which system state one is dealing with.  In field theories with soliton condensates, such as QCD, it also happens that there is a vast number of disordered configurations of about the same action contributing to the  unique vacuum state, and it becomes useful, with a slight abuse of notation, to speak of the entropy of the vacuum as a measure of the number of configurations of these disordered states.  

QCD, like any many-body system, has entropy at finite temperature, where there are a great many pure states that contribute to the mixed thermal state.  Although there are non-trivial problems in evaluating QCD finite-temperature entropy, having to do with infrared singularities and strong coupling, we will not discuss it further except to mention an analogy between thermodynamic entropy and the kinds of zero-temperature QCD entropy of interest to us.  The canonical partition function $\mathcal{Z}_N$ for an isolated thermal system at  fixed particle number $N$ is the trace over all $N$-particle states:
\begin{equation}
\label{thermz}
\mathcal{Z}_N ={\rm Tr} e^{-\beta H}
\end{equation}
Here $\beta = 1/T$ is the inverse temperature and the Hamiltonian is $H$.  This partition function can also be written as an integral over energy, with an entropic weight factor, yielding the Helmholtz free energy:
\begin{equation}
\label{thermzs}
\mathcal{Z}_N=   \int\!\mathrm{d}E\,e^{S(E,N)-\beta E}\approx e^{-\beta (U-TS)}.
\end{equation}
The thermodynamic quantity $U$   is the energy $E$ at the sharp maximum of the integrand, where $\partial S/\partial U = 1/T$.

 Throughout this paper we mostly use the dilute gas approximation (DGA) to describe soliton condensates of QCD.     In the $N$-soliton sector of the DGA grand partition function $\mathcal{Z}$  the action of $N$ solitons is the sum of $N$ single-soliton actions $I_c$   plus a quadratic term whose path integral yields a product $\mathcal{D}$ of determinants times integrals over the collective coordinates $q_i$ arising from  zero modes---derivatives of the $\phi_{ci}$ with respective to the $q_i$--- that  produce zero eigenvalues in some of the determinants.
Then
\begin{equation}
\mathcal{Z}=\sum \frac{1}{N!}\sum_{cc}\mathcal{D}e^{-\sum_i I_c + \int J\sum_i A_{ci}}
\label{zcc} 
\end{equation} 
where the sum over $N$ is the sum over distinct solitons, $cc$ indicates the sum over zero modes, and we suppress writing the determinantal factors.  Note that all solitons have the same action, which needs no index $i$.  The terms $(1/N!)\sum_{cc}\mathcal{D}$ can be regrouped into an entropy factor, much as in Eq.~(\ref{thermzs}).
   In gauge theories such as QCD the action is $\sim 1/g^2$ so that condensates form when the coupling $g^2$ is large enough.  There should be no non-perturbative condensates for a weakly coupled theory, such as electroweak theory.   By comparison with Eq.~(\ref{thermzs}) $g^2$ is analogous to $T$,  and the formation of a QCD soliton condensate at $T=0$ is analogous to a high-$T$ condensed-matter condensate such as a condensate of dislocations at crystal melting.

\section{Outline of the review}

In this review we focus on four QCD entropies at zero temperature, all of which are closely related.  
\begin{itemize}
\item  Soliton entropy:  Of greatest interest are the quantum solitons of an IR-effective QCD action (Sec.~\ref{entsec}) that describes generation of a dynamical gluon mass.  There are center vortices and nexuses in $d=2,3,4$, having codimension 2,  that are important for confinement and chiral symmetry breakdown (CSB).  There are also the usual classical solitons, self-dual instantons and sphalerons, at short distances.
\item Condensate entropy:  QCD has condensates such as $\langle TrG_{\mu\nu}^2\rangle$ and $\langle \bar{q}q\rangle$.  These are global descriptions of soliton condensates.  
\item Entanglement entropy:  Even a pure state, such as the QCD vacuum state, can lead to entropy if the Hilbert space is divided into two parts A and B with the pure state entangled between them, and a partial density matrix for part A is constructed by tracing over the observables in B.  
\item World-line entropy and CSB:  Nearly massless would-be Goldstone particles, such as the pion, have Wilson loops that are highly branched and strongly fluctuating, because of entropy.  Because the  entropy gives negative contributions to the action and  mass of a particle it is critical for  generating a massless pion when only (positive) confining effects are considered.  Entropy also plays a critical role in  generating the $\langle \bar{q}q\rangle$ condensate, both of which accompany CSB.
\end{itemize}

We have no space here to review the fundamentals of non-perturbative QCD that underpin these entropic studies; a few essential points are made in Sec.~\ref{entsec}.  The reader should consult recent books\cite{jgreen,cornbinpap} for details of, {\em e.g.}, the center vortex model and many references.

  Sec.~\ref{genarg} gives some general arguments for the entropy of condensates, not requiring detailed knowledge of their soliton content.   
 Sec.~\ref{solent} reviews the standard lore for estimating the entropy of point-like solitons such as instantons, sphalerons, and $d=2$ vortices.  We go on to more complex estimations for    center vortices in $d=3$ that are analogous to closed polymers.\footnote{The same principles apply to nexuses in $d=4$ (Nexuses are  QCD  magnetic monopoles whose flux tubes are   embedded in center vortex surfaces.  Nexuses disorient center vortex surfaces and give rise to topological charge when a nexus world line links to a vortex surface.} A condensate of closed polymers maps onto a scalar field theory, and entropic effects contribute a negative term to the squared scalar mass.  If there is entropy domination it becomes necessary to add a repulsive $\phi^4$ term for stability.  In $d=4$ center vortices are closed two-surfaces and described by an as-yet unknown string theory; we do not speculate on what such a theory might be like here.  
Sec.~\ref{entconf} reviews the canonical picture of confinement by center vortices, where the area law for a planar Wilson loop arises from an entropic calculation quite similar to that of the entropic rubber band.  However, it is far from clear how to go on to more complex situations, such as a non-planar Wilson loop.  In Sec.~\ref{piercelink} we introduce a crucial   ingredient missing in the techniques of Sec.~\ref{entconf} to show how for a non-planar Wilson loop the area law involves the area of a minimal spanning surface.  The crucial ingredient is the difference between the piercing of any spanning surface by a center vortex and the actual topological linking of the vortex and the surface.  

The techniques of Sec.~\ref{piercelink} have evident analogies to a genuine entropy in the QCD vacuum: Entanglement entropy.      It arises even for a pure state  as the residual entropy of this pure state when it   is entangled between two subsystems A and B, and B (for example, the interior of a black hole) is unobserved. One of the most interesting properties of entanglement entropy is that for many systems, including black holes, entanglement entropy agrees with the holographic principle:  It does not scale with the volume of the black hole, but with its horizon area (as does the Bekenstein-Hawking entropy).   For QCD the pure state is the vacuum, and the unobserved regions will depend on the questions we are asking.      In Sec.~\ref{ent} we briefly review the fundamentals of entanglement entropy and go on to
 point out their relevance to the ideas of Sec.~\ref{piercelink}.

Finally, in Sec.~\ref{csbent} we discuss   the configurational entropy of massless quark world lines and their associated Wilson loops, which     are important in how CSB and confinement are connected.   For massless quarks this entropy contributes a significant negative term to the pion mass, making it possible to be a Goldstone boson with positive area-law and kinetic contributions.  Entropic effects also drive the formation of a $\bar{q}q$ condensate.

No discussion of QCD entropy would be complete without mentioning condensed matter analogs.
There are lattice systems whose ground states are not unique, so their entropy does not vanish at $T=0$.  These include frustrated antiferromagnets,\cite{wann}   water ice,\cite{paul} and  spin ices.\cite{castel,morris}  They provide useful   analogs to the QCD vacuum even though their idealized ground states are not unique. Disorder at $T=0$ arises because there is no energy cost to move from one realization of the system to another.  For example, in a triangular antiferromagnetic lattice once any two spins in a unit cell are oppositely-aligned, the third spin may be anti-aligned with either, but not both, of the first two with the same energy for either anti-alignment.  Spin ices actually have {\em ferromagnetic} interactions obeying the so-called ice rule, calling for two of the four spins in the tetrahedral unit cell to point in, while the other two point out.  Even at zero temperature there is no unique way of arranging  spins to obey the ice rule, hence disorder and entropy.  It takes an extremely small energy to violate the ice rule ({\em e.g.}, three in, one out) in one tetrahedron; a neighboring tetrahedron then flips a spin to obey the ice rule. This spin flip propagates along configurations that have  disorder greatly resembling that of QCD, with quasi-particle magnetic monopoles and Dirac magnetic flux tubes joining them.  Ordinarily a flux tube would cost a fair energy per unit length to form, but not for the spin ice.  Although the physics is quite different, this is analogous to the balancing of entropy and action discussed in  Sec.~\ref{entsec}.

\section{\label{entsec} The IR-effective action for QCD and center vortices}

In $d$-dimensional non-perturbative QCD with gauge group $SU(N)$, the dominant solitons  appear in a gauge-invariant effective action\cite{cornbinpap} that summarizes the fundamental infrared quantum effect of dynamical gluon mass generation:\cite{corn076}
\begin{equation}
\label{effact}
I=\frac{1}{g^2}\int\! \mathrm{d}^dx\, \{\frac{1}{4}G\cdot G+\mathrm{Tr}\frac{m^2}{2}(U^{-1}D_{\mu}U)^2\};\;G\cdot G \equiv \sum G_{\mu\nu a}^2.
\end{equation} 
Throughout this paper we define the gauge potential and field as the canonical ones multiplied by $g$.  Here the trace is normalized to the sum over space-time and group indices,  $D_{\mu}$ is the covariant derivative, and path integrals are indicated by enclosing the differentials in parentheses.\footnote{ We always work in Euclidean space at zero temperature, and suppress writing gauge-fixing and Faddeev-Popov terms.}  The traceless $N\times N$ matrix $U$ transforms according to the adjoint of the gauge group under gauge transformations of the potential.  The running coupling $g^2$ and mass $m^2$ are each evaluated at zero momentum.  

This effective action has a number of solitons, among them center vortices with co-dimension 2.
Vortex-like excitations, and the importance of entropy dominance, were invoked before center vortices\cite{bmk} through simple Abelian lattice models.  For non-Abelian gauge theories in $d=3,4$,      't  Hooft\cite{thooft} introduced center vortices that were soon thereafter interpreted\cite{corn066}    as solitons of the effective action of Eq.~(\ref{effact}).
We will be explicit only for   $d=3$, where a simple center-vortex soliton has the form:
\begin{equation}
\label{cvsoliton}
A_i(x;j)=2\pi Q_j\epsilon_{iab}\partial_a \oint_{CV}\!\mathrm{d}z_b[\Delta_0(x-z)-\Delta_m(x-z)].
\end{equation}
The diagonal matrices $Q_j,\;j=1\dots N-1$ are such that $\exp [2\pi i Q_j]=\exp [2\pi i/N]$ is in the center of the group, and as before $\Delta_m$ is the free propagator of mass $m$.  In other dimensions the integral is over a closed surface of codimension 2, and the epsilon symbol has $d$ indices.  This simplest of center vortices is essentially Abelian, but there are many ramifications, such as nexuses that disorient the vortex and contribute to fractional topological charge, that are essentially non-Abelian;\cite{cornbinpap} we have no space to study them here.  Note that there is no singularity in the vortex on its defining contour $CV$, because short-distance singularities cancel between the two propagators.   In any dimension center vortices have finite transverse thickness $1/m$ coming from the massive propagator.    The term in $\Delta_0$, considered by itself, is a pure (singular) gauge transformation of long range; it alone is responsible for confinement.

The dynamical gluon mass, absent at the Lagrangian level, is a consequence of condensates such as center vortices and nexuses.  If there are no condensates there can be no gluon mass, since it has been shown\cite{lavelle} that at large momentum this mass runs as:
\begin{equation}
\label{runmass}
m^2(q^2)\sim \frac{\langle G\cdot G \rangle }{q^2}
\end{equation}
(just as the CSB running quark mass runs like $\langle \bar{q}q\rangle/q^2$).  And, of course, if there is no mass scale in the effective action there can be no solitons such as center vortices and nexuses.  So the gluon mass, on which all non-perturbative effects depend, is entropy-driven.

A major concern of this review is the area-law behavior of Wilson loops coming from their coupling to a condensate of randomly-distributed center vortices.  We evaluate this for a single vortex and a Wilson loop $\Gamma$, all of whose lengths are large compared to $1/m$.  The massive propagator in Eq.~(\ref{cvsoliton}) yields only perimeter terms and can be dropped, yielding:
\begin{equation}
\label{wleval}
W\equiv \mathrm{Tr}Pe^{i\oint_{\Gamma}\!dx_iA_i(x)}   = \exp [\frac{2\pi i}{N} \oint_{\Gamma}\!\mathrm{d}x_i\!\oint_{CV}\mathrm{d}x_j\epsilon_{ijk}\partial_k\Delta_0(x-z)]
\end{equation}
where $\Gamma$ is the Wilson loop contour and $CV$ labels a given center vortex loop.  The coefficient of $2\pi i/N$ in the integral on the right-hand side is a topological quantity, the Gauss linking number,\footnote{This is true in all dimensions.} so for a single vortex 
\begin{equation}
\label{onevort}
W=e^{2\pi i L_k/N}.
\end{equation}
The integral over the loop contour can be rewritten with Stokes' theorem to become an intersection number of the center vortex with any surface that spans the Wilson loop.  This raises interesting questions about which area appears in the expectation value $\langle W\rangle$, which will be taken up in Sec.~\ref{piercelink}.

The next question is how to describe the entropy of a condensate of center vortices or other solitons.

\section{\label{genarg} General arguments for condensate entropy}

 Is it possible to say anything about the entropy of a condensate, without even knowing what the condensate is made of?  Perhaps surprisingly, the answer is yes.  Consider first an often-discussed candidate for a ``condensate" that has zero entropy:  A covariantly-constant chromomagnetic field in $d=4$.  Use the notation
\begin{equation}
\label{thetanot}
\theta (x) =  \frac{b}{8}G\cdot G
\end{equation}
where $b=11N/48\pi^2$ is the no-quark one-loop coefficient in the beta-function $\beta (g) =-bg^3$ for $SU(N)$.
The one-loop action   can be found\cite{savvidy} by the standard techniques of Schwinger, and yields for the real part:
\begin{equation}
\label{effpot}
\Gamma (\theta )=\int \!\mathrm{d}x\,\theta \ln \{ \frac{\theta}{e\langle \theta \rangle}\}.
\end{equation}
This is an interesting expression because it has a minimum of {\em negative} value $-\int \langle \theta \rangle$ at $\theta = \langle \theta \rangle$, suggesting the dominance of entropy over (positive) action.
Unfortunately, it has been pointed out\cite{nielol} that, just as in the Schwinger calculation of vacuum decay for a constant electric field in QED, one should add to this action an imaginary part, signaling instability of a constant chromomagnetic field in QCD.  Nielsen and Olesen\cite{nielol} showed that the instability could be removed with sufficient space-time inhomogeneity, and correctly interpreted this as a signal for formation of high-entropy disordered domains.  

But one need not assume a covariantly-constant field to arrive at $\Gamma$ of Eq.~(\ref{effpot}).   We can find this action, with no imaginary part,  using methods based on the one-loop renormalization group.\cite{pagtom,nsvz}  Of particular interest is an infinite set of relations\cite{nsvz} among zero-momentum components of $\theta$:
\begin{equation}
\label{nsvz}
\int \dots \int \![\prod^N\mathrm{d}x_i]\langle T\theta (x_1)\dots \theta (x_N)\theta (0)\rangle_{conn} = \langle \theta (0) \rangle
\end{equation}
found by repeatedly applying $g^3\partial /\partial g$ to   the vacuum energy density $\epsilon_{vac}=(1/4)\langle T^{\mu}_{\mu}\rangle$ as defined by the partition function:
\begin{equation}
\label{zerosource}
\mathcal{Z}(0)\equiv e^{-\int\!\mathrm{d}x \,\epsilon_{vac}}=\int\!(\mathrm{d}A_{\mu a})e^{-\frac{1}{4g^2}\int G\cdot G} 
\end{equation}
from which it follows that
\begin{equation}
\label{evac}
g^3\frac{\partial}{\partial g}\epsilon_{vac}=\frac{-1}{2}\langle G\cdot G \rangle.
\end{equation}
Because $\epsilon_{vac}$ is renormalization-group invariant of dimension 4 it satisfies
\begin{equation}
\label{rginv}
(\mu \frac{\partial}{\partial \mu} +\beta (g)\frac{\partial}{\partial g})\epsilon_{vac}= (4+\beta (g)\frac{\partial}{\partial g})\epsilon_{vac}=0,
\end{equation}
which, with Eq.~(\ref{evac}), yields the trace anomaly.  Repeated differentiation of Eq.~(\ref{zerosource}) then yields the sum rules of Eq.~(\ref{nsvz}).

At this point we have made no use of the putative one-loop effective action of Eq.~(\ref{effpot}).   But standard Legendre transform methods show\cite{corn082} that $\Gamma (\theta )$ in Eq.~(\ref{effpot}), {\em with no imaginary part}, is precisely the effective action which gives rise to these sum rules.  Moreover, the generating functional found by exponentiating Eq.~(\ref{effpot}) is\cite{corn109} a Poisson distribution for $\theta$, if this quantity is thought of as the sum of a large number of independent random terms with average value $\langle \theta \rangle$, so that $\theta = \sum \theta_i$.  This suggests strongly that the $\langle G\cdot G\rangle$ condensate has a large entropy, arising from lumps of condensate with a finite correlation length.

These considerations depend critically on using the beta-function at one loop only.  But in $d=3$ (where there is no beta-function) the infinite sum rules analogous to those  of Eq.~(\ref{nsvz}) and the corresponding effective action are exact, given the reasonable hypothesis that $d=3$ QCD has only one mass scale, $g^2$ itself.\cite{corn109}  In $d=3$ we define $\theta$ as:
\begin{equation}
\label{d3theta}
\theta (x)=\frac{1}{4g^2}G\cdot G,
\end{equation}
an operator with mass dimension 3.  Because $g^2$ has mass dimension 1, $\epsilon_{vac}$ necessarily scales like $g^6$ in the absence of any other mass scale.  This might seem impossible since in perturbation theory  high-order graphs have high powers of $g$.  But all graphs of eighth or higher order are UV finite; their IR divergences are removed by a gluon mass $m\sim g^2$, so ordinary dimensional counting works.  Eqns.~(\ref{zerosource},\ref{evac}) continue to hold in $d=3$, and combined with $\epsilon_{vac}=Kg^6$ for some pure number $K$ yields:
\begin{equation}
\label{evac2}
\epsilon_{vac}=-\frac{1}{3}\langle \theta \rangle.
\end{equation}
Since $\langle \theta \rangle$ is positive, the vacuum energy is negative and hence entropy-dominated.\footnote{This equation is actually a simple consequence\cite{corn109} of Lorentz invariance and isotropy of the vacuum in $d=3$.}  
As in $d=4$, continued application of $g^3\partial /\partial g$ to $\epsilon_{vac}$ yields an infinite set of sum rules:
\begin{equation}
\label{d3sumrule}
\int \dots \int \![\prod^N\mathrm{d}x_i]\langle T\theta (x_1)\dots \theta (x_N)\theta (0)\rangle_{conn} = \frac{(N+2)!}{3!}\langle \theta (0) \rangle.
\end{equation}
Again, standard Legendre transform techniques yield the effective action:
\begin{equation}
\label{d3effact}
\Gamma (\theta)=\int\!\mathrm{d}^3x\,(\theta -\frac{4}{3}\theta^{3/4}\langle \theta \rangle^{1/4}).
\end{equation}
This has its minimum value of $-(1/3)\langle \theta \rangle$, as required.  

  There is\cite{corn114} a complex scalar field  theory with a $|\psi |^4$ interaction  in which    loop corrections yield a vacuum energy of   precisely the  form of Eq.~(\ref{d3effact}), with the identification $  \theta   \sim   |\psi |^4 $.  The peculiar 3/4-power term comes from loop corrections yielding (specific to $d=3$)  $|\psi |^3$ terms that, it turns out, are equivalent to a negative squared mass.   Just such a theory is needed to describe a $d=3$  center-vortex condensate, as we review in Sec.~\ref{d3cent}.

\section{\label{solent} Entropy of solitons}

\subsection{\label{pointsol} Entropy of instantons}

The simplest expression of entropy is for point-like solitons, such as instantons.  Strictly classical instantons, the solitons of the usual ($m=0$) action, can have any size.  But in the IR-effective action these are approximately solitons only if their size $\rho$ is very small in the sense $m\rho \ll 1$.  Otherwise, the solitons analogous to instantons (which are no longer self-dual) have size limited by $1/m$.  Similarly, the barrier height represented by sphalerons cannot vanish, but is of order $m$ at least.  For simplicity, and because we are interested only in IR effects, we will proceed as if instantons were of fixed size and ignore the size collective coordinate.  

Set the currents $J$ to zero in the generating functional equation (\ref{zcc}), since they do not contribute to the entropy, and add a chemical potential term $\exp [\mu N]$ in the summand over $N$.    This generating functional at zero current is just the grand partition function at zero temperature.     The only integral over collective coordinates that we need indicate explicitly yields a factor $V_4/V_c$ for every instanton, where $V_c$ is a finite volume (depending on $g^2$)to be calculated from the determinants.   Then, using Stirling's formula and replacing the sum over $N$ with evaluation of the summand at $N=\bar{N}\sim V_4$ we find
\begin{equation}
\label{partfunct3}
\ln \mathcal{Z}=\bar{N}[\ln (\frac{eV_4}{\bar{N}V_c})-I_c+\mu]
\end{equation}
This is equivalent to  the standard thermodynamic result $\ln \mathcal{Z}=\beta [-U+TS+\mu \bar{N}]$ for free particles with internal energy, and so we can identify the entropy per instanton as the logarithmic term  in the brackets, and the free internal energy per particle, in units of temperature $T$, with the classical action $I_c$.   

It was once argued that instantons (and by extension, other solitons) could not contribute much at large $N_c$ for color gauge group $SU(N_c)$, because of 't Hooft scaling $g^2\sim 1/N_c$ and the action factor is suppressed exponentially in $N_c$.  But in fact the {\em leading} term in the collective coordinate integral exactly cancels this action suppression.\cite{grossmat}  What happens at non-leading orders is unknown, but there is no known reason to believe that instantons are suppressed at large $N_c$.  The same leading-order exact cancellation between action and entropy also takes place for center vortices, at least in $d=2$, where they are point particles.\cite{corn121}

More elaborate versions of the above argument were used early on to argue for confinement by instanton condensates, but these ultimately failed for a number of reasons that we will not go into here.  Perhaps it should be noted that recent lattice simulations suggest that dilute instantons and anti-instantons do not constitute a large fraction of the QCD vacuum configurations.\cite{alexhor}  

\subsection{\label{d3cent}  Entropy of center vortices}

Dealing with the configurational entropy of center vortices is conceptually straightforward in $d=3$, where a condensate of oriented closed strings can be mapped onto a complex scalar field theory (a trick long used by polymer physicists).  The idea\cite{stonethom} is that the partition function for a condensate of these strings is exactly the same as the partition function for the free  complex scalar  field theory, once one recognizes that the propagators in the vacuum graphs  of the scalar field theory can be written in Schwinger-Feynman proper-time form as an integral over closed strings.  First, the standard partition function for a free complex scalar theory of mass $M$ is
\begin{eqnarray}
\label{standpart}
\mathcal{Z} & = & \int\,(\mathrm{d}\psi \mathrm{d}\bar{\psi})e^{-I_0}=\exp [\mathrm{Tr}\ln \Delta_M]  \\  \nonumber
I_0 & = & \int\!\mathrm{d}^3x\,|\partial \psi|^2+M^2|\psi |^2 
\end{eqnarray}
where $\Delta_M(x,y)$ is the free propagator, and the Tr is over all Euclidean space-time (and possibly other discrete degrees of freedom).
Up to a normalization factor the trace of the logarithm of a scalar propagator (in any dimension) has the proper-time representation as a path integral over closed world lines:
\begin{equation}
\label{proptime}
{\rm Tr}\ln [\Delta_M]=\int_0^{\infty}\frac{ds}{s}\oint(dz)\exp \{-\int_0^sd\tau [\frac{1}{2}\dot{z}^2+\frac{M^2 }{2}]\}.
\end{equation}
Now note that in $d=3$ the probability $P(x,x';t)$ that a random string made up of $t$ steps, of average length $a$, will go from $x$ to $x'$ has the continuum random-walk path-integral form 
\begin{equation}
\label{probint}
P(x,x';t) =\int_x^{x'}\,(\mathrm{D}z)\exp [-\int_0^t\,\mathrm{d}\tau\frac{3\dot{z}^2}{2a}]
\end{equation}
(up to a normalization) where $\mathrm{d}\tau\approx a$.  Here $a$ represents the correlation length for the string to change directions by, say, $\pi /2$.

The crucial step in estimating the entropy is to find the number of ways to go from $x$ to $x'$ with a string of length $t$.  Imagine that the string lies on the links of a cubic lattic of side $a$.  The string grows by addition of a new link at one end, which can be done in $2d-1=5$ ways when backtracking is disallowed.  Then the number of ways to go from $x$ to $x'$ is approximately $P(x,x';t)$ times $\exp [5\ln t]$.  We weight each link of the string by a factor $\exp [-\mu ]$, or $\exp [-\mu t]$ for the whole string; for the field theory propagator this represents the mass factor in Eq.~(\ref{proptime}).  

The weighted total number of distinguishable ways a single {\em oriented} closed string going from $x$ to $x$ can be formed is roughly
\begin{equation}
\label{onestring}
5^t\int_0^{\infty}\,\mathrm{d}\tau \frac{1}{\tau } P(x,x;\tau )e^{-\mu t}\equiv \mathcal{Z}_1
\end{equation}
which defines the one-string partition function $\mathcal{Z}_1$.
The factor $1/\tau $ corrects for double counting:  Divide by the length of the string, because any point on a given string serves to label a single configuration.    The grand partition function is, as usual, 
\begin{equation}
\label{grandstring}
\mathcal{Z}_{string}=\sum\frac{[\mathcal{Z}_i]^N}{N!}=\exp \mathcal{Z}_1.
\end{equation} 

We are almost done.  Change variables via $\mathrm{d}s=(a/3)\mathrm{d}\tau$, and then define
\begin{equation}
\label{massdef}
M^2=[\frac{3}{a^2}][\mu -\ln 5].
\end{equation}
The resulting grand partition function $\mathcal{Z}_{string}$ is precisely the free scalar partition function of Eq.~(\ref{standpart}).  Observe that entropy can dominate over the standard mass term, in which case   $M^2$ is negative, and the path integrals defining the scalar propagators make no sense.  This is, of course, just the way the mass term behaves in spontaneously-broken theories.  Such theories have to be stabilized by a $|\psi |^4$ interaction, and then there is a phase of the theory in which $\langle |\psi | \rangle$ is a non-vanishing constant throughout space-time, signaling a condensate.  In our case we interpret the $|\psi |^4$ interaction as mimicking the mutual avoidance of different strings, since the overlap action of two string trying to occupy the same space-time amounts to a repulsion.   The condensate forms because the entropy per unit vortex length dominates the action per unit length.  Then vortices tend to become very long; in fact,\cite{stonethom}  a finite fraction of them have infinite length.   

From the explicit form of the $d=3$ center vortex in Eq.~(\ref{cvsoliton}) the expectation value $\langle W\rangle$ of a Wilson loop $\Gamma$ is equivalent to coupling a gauge potential to the field $\psi$.  For the simple case of $SU(2)$, where there is only one matrix $Q_j$ with entries $\pm 1/2$,   the vortex is effectively Abelian with $Q_j=1/2$ a number and not a matrix, because every configuration is accompanied by an equal contribution from its complex conjugate.   This vortex is:
\begin{equation}
\label{vpot}
V_i(y)=\pi \oint_{\Gamma}\!\mathrm{d}x_j\epsilon_{ijk}\partial_k[\Delta_0(x-z)-\Delta_m(x-z)].
\end{equation}
If one is only interested in the area-law part of the Wilson loop, the $\Delta_m$ propagator of Eq.~(\ref{cvsoliton}) can be dropped.

This model's implications for confinement and for Chern-Simons fluctuations have been explored\cite{corn114}, beginning with  a complex scalar field theory with $|\psi|^4$ coupling. Quantum corrections yield the results of Sec.~\ref{genarg} for entropy generation in $d=3$.  These considerations have been generalized\cite{corn133} to non-Abelian gauge theories beginning from the center vortex solitons of Eq.~(\ref{cvsoliton}) and adding   nexuses.   The partition function of the IR limit of the gauge theory  has the form of the string partition function of Eqs.~(\ref{onestring},\ref{grandstring}) after a simple redefinition of variables.   

Four dimensions is much more complicated, since the center-vortex partition function naively maps onto a string theory in a non-critical dimension.   It will be essential to modify the action of this string theory with curvature-dependent terms accounting for stiffness and bending resistance of the vortex surfaces, coming from increases in the fundamental gauge-theory action when a vortex surface is bent.   However, these complications are not particularly important for understanding the basics of confinement, which we present in the $d=3$ context.
 
\section{\label{entconf}  Entropy and confinement}

The Wilson-loop area law is actually entropic.  As remarked in Sec.~\ref{entsec}, a single center vortex contributes a phase factor
\begin{equation}
\label{wiloop}
W=P\exp [i\oint_{\Gamma} \!\mathrm{d}x_iA_i]=\exp [\frac{2\pi iL_k}{N}]
\end{equation}
where $L_k$ is the Gauss linking number of the vortex tube and the curve $\Gamma$.  The phase factor for many vortices is the product of the phase factors of the individual vortices, which do not interact.  For simplicity consider only $SU(2)$, where the fundamental-representation Wilson loop for a single vortex is $(-1)^{L_k}$.  Even Gauss link numbers contribute nothing to confinement; only odd links count.

\begin{figure}
\begin{center}
\includegraphics[width=3in]{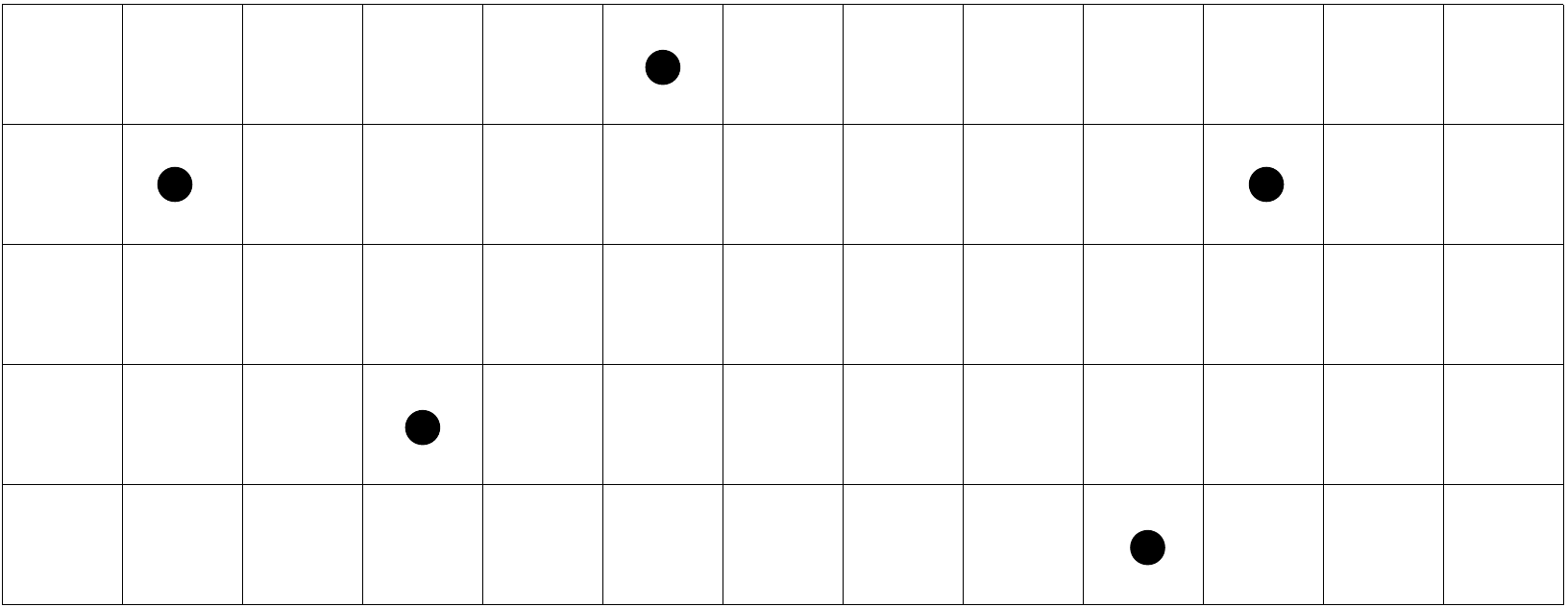}
\caption{\label{loop} A rectangular Wilson loop whose flat spanning surface is divided into $N$ squares whose length is the correlation length $\lambda$.  Either zero or one vortex can pierce any square (black dots indicate a piercing}.  
\end{center}
\end{figure}

To find\cite{jgreen,cornbinpap} the area law of $\langle W\rangle$ choose a simple loop, consisting of a rectangle spanned by a flat surface (see Fig.~
\ref{loop}).   
The spanning surface is divided into $N$ squares whose side is the correlation length $\lambda$, so that the area $\mathcal{A}$ of the square is $\mathcal{A}=N\lambda^2$.  This length is (roughly) defined by saying that no more than one vortex can pierce, or intersect, such a square, so a square can either be singly-occupied (with probability $p$) or empty (with probability $1-p$).  Because we use the DGA, we take it that $p\ll1$.

The black dots indicate where a vortex has pierced the Wilson-loop surface; we assume such piercings are random and independent of one another.  Although a piercing is not  necessarily the same as a linkage  (that is, $L_k$ is odd) to the loop, because it simplifies things so much we assume that every piercing indicates a link number of unity.  In that case, the expectation value $\langle W \rangle$ is just finding the probability $P_N(J)$ that there are $J$ linkages, so defined, and calculating the expectation value of $(-1)^J$ with this probability.  

Here is the first place where entropy enters, and it enters in precisely the same way as it does for the entropic rubber band.  By simple combinatorics the probability $P_N(J)$ is
\begin{equation}
\label{comb}
P_N(J)=\frac{N!}{J!(N-J)!}p^J(1-p)^{N-J},
\end{equation}
the same as for an anisotropic rubber band, where the probability $p$ of a left-pointing link is less than that of a right-pointing link.  The factorial factors are just the number of ways of distributing $J$ things in $N$ slots.  To relate this probability density to an entropy density, note that there are a total of $2^N$ states, with an entropy of $N\ln 2$.  The number of states for a given occupancy number $J$ is the product $2^NP_N(J)$, whose logarithm defines a partial entropy at fixed $J$.  In expectation values this $2^N$ factor cancels with the same factor in the partition function.

 We easily calculate
\begin{equation}
\label{wloopexp}
\langle W \rangle = \sum_J\,(-1)^JP_N(J) = (1-2p)^N=e^{-K_F \mathcal{A}}
\end{equation}
where 
\begin{equation}
\label{sigma}
K_F = -\frac{\ln (1-2p)}{\lambda^2}\approx \frac{2p}{\lambda^2}\equiv 2\rho_2
\end{equation}
and $\rho_2$ is the areal density of pierce points.  [Because the codimension of a vortex is 2, vortices are characterized in any dimension by an areal density of pierce points, and a confining Wilson loop is  an area law in any dimension.]  Note that the areal density is necessarily the same anywhere in spacetime.

However enticing the simplicity of this ``derivation" of an area law may be, it has a serious flaw, which when resolved will reduce the entropy:  The piercings are correlated.  It is much more complicated to deal with this issue, which as it turns out does not interfere with the basic fact that confinement means an area law.\cite{corn135}  Furthermore, the derivation does not explain the ``obvious" point that the surface whose area appears in the area law  is a {\em minimal} spanning surface.  How does one show that for a non-planar contour?

\section{\label{piercelink} The difference between piercing and linking}

One should not conflate piercing with linking.  There are critical circumstances in which a vortex necessarily pierces a spanning surface an {\em even} number of times, in which case it makes no contribution to the area law. Nonetheless even-numbered piercings are important, because they render inactive portions of the spanning surface that might otherwise be linked.  This inactivation changes the entropy crucially for non-minimal spanning surfaces.  (There is also a certain amount of inactivation even for a minimal surface, but it is a quantitatively non-leading effect.\cite{corn135}) 

\begin{figure}
\begin{center}
\includegraphics[width=3in]{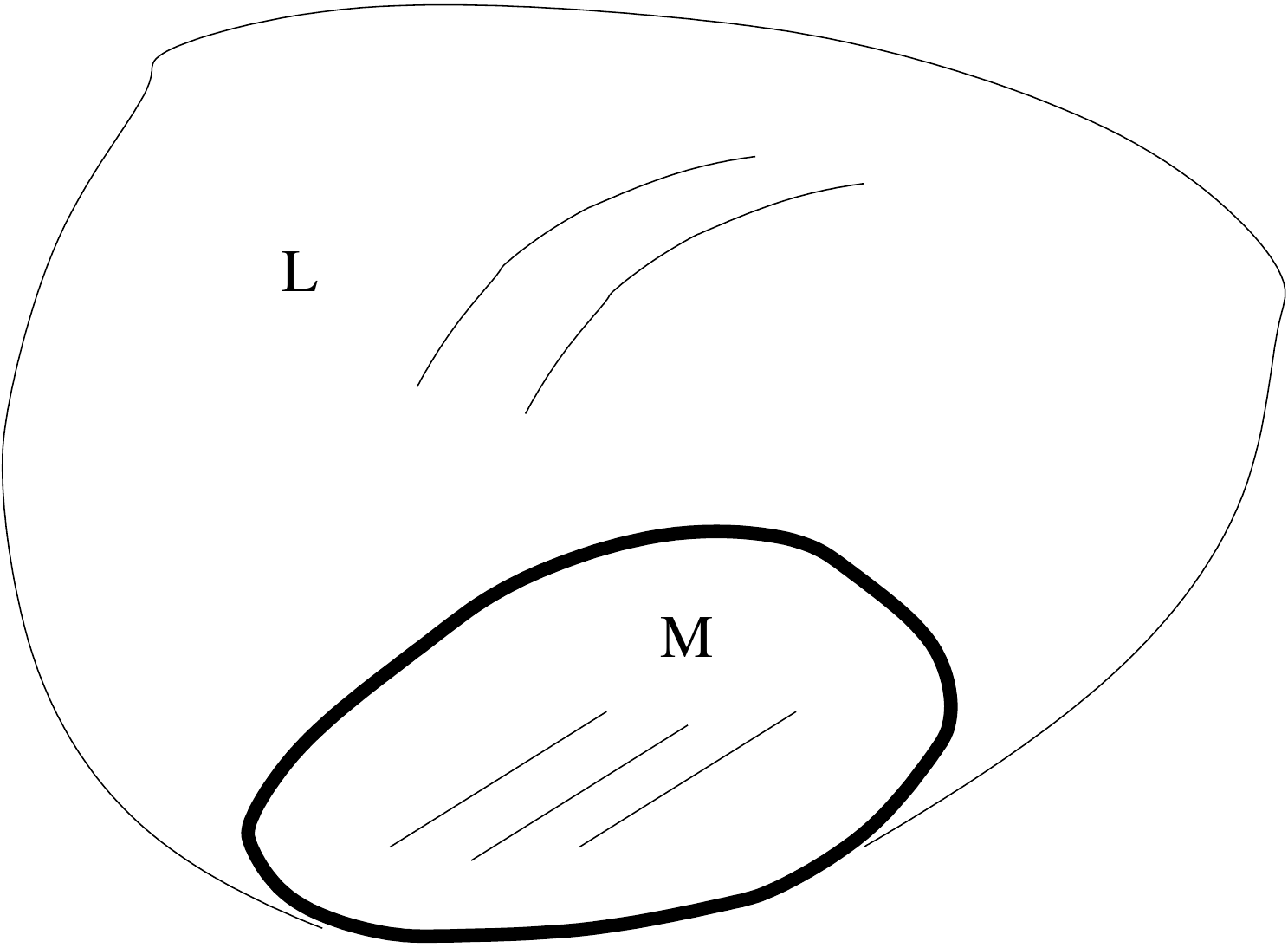}
\caption{ \label{surfloop} Two surfaces spanning a non-planar Wilson loop (heavy line):  A minimal surface M, and a much larger  surface L.  How to tell which area, M or L or either, to use in the area law?  }
\end{center}
\end{figure}

 Draw a generic ({\i.e.}, non-planar) large Wilson loop such that it has a unique  minimal spanning surface M, and then erect a different spanning surface L with necessarily greater area so as to form a closed surface on which the Wilson loop lies (see Fig.~(\ref{surfloop})).  The areas of these surfaces are $\mathcal{A}_M,\mathcal{A}_L$.   (For reference to Sec.~\ref{ent} on entanglement entropy, we now have a closed surface M+L dividing three-space into two parts A and B, with B the region interior to the surface M+L.)   Which area are we to use in the Wilson loop?
Since the contribution of a given vortex to $\langle W\rangle$ is a purely topological signed intersection number, it is the same, and more important it is also the same (mod 2),  for either spanning surface M or L.

Is there a unique answer for the area law coming from vortices?      The answer is yes, and explains why the area law involves a minimal spanning surface.   The demonstration only uses knowledge of what happens on the surfaces M and L.
 Only vortices entering  the volume B by piercing  M and exiting by piercing L (or {\em vice versa}) give  a linkage with the Wilson loop.  There are, by assumption, $N_M\equiv \rho_2\mathcal{A}_M$ vortices piercing M.  By the same assumption $N_L\equiv \rho_2\mathcal{A}_L$ vortices pierce L, and $N_L>N_M$.  In fact, $N_M$ is the least number of vortices (up to small fluctuations) that can pierce any spanning surface.  It should therefore be clear that there are $(N_L-N_M)/2$ vortices piercing surface L that do not pierce surface M at all; they exit through surface L again. Their only role is to render inactive enough vortex sites so that exactly $N_M$ are really linked.  So in using surface L to calculate the area law by invoking Eq.~(\ref{wloopexp}), the number of vortices entering this formula is $N_M$, not $N_L$, so we find the unique minimal area law.   A study\cite{corn135} of the  question glossed over here, of the quantitative difference between the number of vortices piercing a flat surface and the number linked to it, shows that for dilute vortices the string tension is slightly reduced from its canonical value of $2\rho_2$.

Here is an explicit computation of the area law for a simple non-minimal surface, showing the importance of distinguishing piercing and linking.  Change the spanning surface M to be slightly non-planar by excising one interior square, and erecting a cube of side one unit over this square; the bottom square of the cube is missing. We call this cube the bump. To generalize the entropic calculation of Eq.~(\ref{comb}) use the identity:
\begin{equation}
\label{iden}
\prod^N(A_i+B_i)\equiv \prod B_i+\sum A_i\prod_{j\neq i}B_j+ \sum_{i<j} A_iA_j\prod_{k\neq i,j}B_k +\dots + \prod A_i.
\end{equation}
Suppose that every square in the lattice of Fig.~\ref{loop} has a different probability $p_i$ of piercing and a probability $1-p_i$ of no piercing.  If we identify $A_i=p_i,B_i=1-p_i$, the right-hand side of Eq.~(\ref{iden}) decomposes the total (unit) probability into terms with no, one, two,$\dots$ piercings.  To find $\langle W\rangle$ for this lattice, just set $A_i=-p_i,B_i=1-p_i$.  The result is:
\begin{equation}
\label{newloop}
\langle W\rangle = \prod^N (1-2p_i)\approx e^{-2\sum p_i}
\end{equation}
which, of course, agrees with Eq.~(\ref{wloopexp}) when $p_i\rightarrow p$.  For the lattice with the bump, there are $N-1$ squares with probability $p_i=p$, and 5 squares with  probability $p/5$.  This follows because  if a vortex (which is now allowed to form a three-dimensional closed path along the dual lattice) pierces any of the 5 squares in the bump, there is only one way to have an odd link number, and that is to go out the empty bottom square.  All other exits have even (in fact, zero) linking number.    Using these results in Eq.~(\ref{newloop})  yields exactly the result in Eq.~(\ref{wloopexp}) for the flat lattice without any bump.  If we had just counted piercings, without asking whether they amounted to linkings, this would certainly not be so.  It is more tedious to do these calculations for various other bumps (two adjoining cubes,...), but one always gets the same result:  1) Any two surfaces spanning the flat Wilson loop give the same area law; 2) This area law is that of the minimal surface.

Our use above of the closed surface M+L that bounds a three-volume where we know nothing about what happens to the center vortices in the interior strongly suggests an analogy with entanglement entropy, \cite{bomb,sred,cala,kitpres,ryutak,vel,buipol,eisert}  as pointed out in    Sec.~\ref{ent} immediately below.   As before, the calculation of an actual area law is the same as calculating the entropy of vortex link points on these surfaces.

  In view of the analytic difficulties in distinguishing piercing from linking, it would be worthwhile to do lattice simulations by constructing ensembles of random mutual- and self-avoiding vortices on a lattice dual to the lattice on which Wilson loops and surfaces live, and calculating expectation values of the type in Eq.~(\ref{wloopexp}) numerically for a variety of non-planar Wilson loops,\cite{corn139}  to establish that the area law holds with the minimal area and a universal (loop-independent) string tension.  One can then also study fluctuations in the area of the area law, to establish such effects as the L\"uscher term.  These simulations will be far easier than simulations of full QCD.

\section{\label{ent} Entanglement entropy}

Usually one thinks of the entropy of a pure state $|\psi \rangle$ of a system as being zero.  However, if the system can be decomposed into two subsystems A, B  whose sum is the whole system, and one of the systems is unobservable for some reason (for example, B is the region inside the horizon of a black hole), the partial entropy of the remaining subsystem will not vanish even for that pure state. This can happen when the pure system state is an entangled state of the two subsystems. We define a density matrix and  entropy for the A subsystem in the usual way:
\begin{equation}
\label{entangent}
\rho_A={\rm Tr_B} \,\rho_{A+B} = {\rm Tr_B}|\psi \rangle\langle \psi |,\;\;S_A=-{\rm Tr_A} \,\rho_A \ln \rho_A.
\end{equation}  
Suppose A and B are  two-level systems, and we denote their eigenfunctions in the usual way as $|+\rangle$ and $|-\rangle$.  We calculate the entanglement entropy of the pure but entangled state
\begin{equation}
\label{entstate}
|\psi\rangle = \cos \alpha |+-\rangle + \sin \alpha |-+\rangle
\end{equation}
(state of subsystem A written first in the ket) from
\begin{eqnarray}
\label{entex}
\rho_A & = & \cos^2\alpha |+\rangle \langle +|+ \sin^2 \alpha |-\rangle \langle -|,\\ \nonumber
S_A & = & -\cos^2 \alpha \ln \cos^2\alpha -\sin^2\alpha \ln \sin^2\alpha.
\end{eqnarray}
This vanishes if $\cos \alpha$ or $\sin \alpha$ vanishes, when the state is not entangled.

Entanglement entropy has many applications in condensed matter theory,\cite{cala,kitpres,eisert} field theory,\cite{bomb,sred,buipol,vel} and string theory.\cite{ryutak}     A standard example is a nearly-static black hole in $d=3+1$ space-time whose horizon divides three-space into two regions A (observable)  and B (unobservable).     As is well-known, the entropy of the black hole does not scale with the volume inside the horizon, but with its area.  

We have already seen such a feature arise in QCD in Sec.~\ref{piercelink},    where the closed surface $M+L$  of Fig.~\ref{surfloop} divides three-space into outside (A) and inside (B) regions. The  entanglement entropy scales with the area of region B.  The  observable (that is, in region A) center vortices of the vacuum are closed loops, but  those closed loops entering and leaving the surface of region B appear to be open strings in A, terminating on this surface.   Such loops are, of course, entangled with their continuations in region  B.  Just from observations in region A, one sees the open strings and their endpoints on the surface, very much the same as if the surface were a brane with the chromomagnetic flux tubes  of center vortices terminating on it.\footnote{The identical argument has been given\cite{buipol} but with the  postulation of a disordered condensate of chromoelectric rather than the chromomagnetic flux tubes of the center-vortex condensate.}   We have already calculated the entropy associated with the random positions of these endpoints, which behave for all practical purposes like a gas of point particles on the surface, with the usual vortex areal density $\rho_2$, and found a unique minimal-surface area law.

Another example\cite{corn130} of entanglement entropy (not termed as such) is based on a $d=3$ analog of the original L\"uscher bag\cite{lusch} that divides four-space into an outside A and an inside B.  In this example  the Chern-Simons term plays the role in the three-dimensional bag that the $G_{\mu\nu}\tilde{G}_{\mu\nu}$ density does in the  L\"uscher bag. We interpret this bag as region B, with an unobserved Chern-Simons density.  The volume integral of this density can be expressed as a surface integral that expresses the holographic entanglement entropy.  Entanglement entropy from long-range topological disorder is also important in condensed matter physics.\cite{kitpres}

\section{\label{csbent} Entropy of Wilson loops:  Chiral symmetry breaking}

The idea of chiral symmetry breaking (CSB) goes back to the fifties, well before QCD, yet it is still an active field of research.  Lattice simulations have helped, but not singled out a definitive mechanism.  For $SU(2)$ they show\cite{deforc}  that center vortices not only confine quarks, but are also critical for CSB:  When they are artificially removed, both confinement and CSB are gone. This is consistent with earlier continuum arguments\cite{corn070} that confinement is necessary and sufficient for CSB.  There is also lattice evidence that the   deconfinement transition temperature is rather close to the CSB restoration temperature for quarks.   

For decades people have embraced the idea that one-gluon exchange is strong enough to produce CSB without confinement; we examine the opposite case, where confinement is needed.  Note, however, that a recent study\cite{agpap} concludes that one-gluon exchange is enough for CSB, based on a study of the pinch technique\cite{cornbinpap} Schwinger-Dyson equations based on lattice results and including a gluon mass, which tends to weaken  one-gluon effects.   
\subsection{The importance of entropy for the pion}

Up to now the Wilson loops that entered our study of confinement were very large and prescribed beforehand.  Since they were fixed, they had no entropy at all.  The physical interpretation of such large smooth Wilson loops is that they refer to very massive   particles, with mass $\mathcal{M}\gg K_F^{1/2}$  ($K_F$ is the string tension).  Such particles can be far apart (of order $\mathcal{M}/K_F$)  and hardly feel the confining force.  

We now want to study dynamical Wilson loops in the opposite limit, of massless quarks that make up the massless   Goldstone pion, a $\bar{q}\gamma_5q$ bound state.\cite{corn070,corn140}   We work in the quenched limit and omit quark determinants in path integrals.
    In the quenched limit the propagator of the pion field $\pi (x)$ is:
\begin{equation}
\label{pion}
\langle T\pi (x) \pi (0)\rangle \sim \int\!(\mathrm{d}A)e^{-I(A)}{\rm Tr}[S_A(x,0)\gamma_5S_A(0,x)\gamma_5]
\end{equation}
where $S_A$ is the   quark propagator in the gauge potential $A$; it is an integral over paths, in the same spirit as  Eq.~(\ref{proptime}).  The trace creates a Wilson loop starting at 0, passing through $x$, and returning to 0.  Between these endpoints, the Wilson loop  fluctuates, and if the mass parameter in the propagator is small enough compared to   $K_F^{1/2}$   these fluctuations are large and have large entropy that will make important {\em negative} contributions to the mass of the pion.  Such negative contributions are essential, if one-gluon exchange is (as we assume) too weak to produce CSB by itself.  

The distance $r$ between $q$ and the $\bar{q}$ in the pion cannot be large; if it were, the confinement energy $K_Fr$ would be considerably larger than such scales as the running quark mass at zero momentum $M(0)$.  On the other hand, the $\bar{q}\gamma_5 q$ Wilson loop can propagate a long distance, because the pion it represents is massless.  So schematically we can think of the pion Wilson loop as very long and thin, and apply the ideas used earlier in Sec.~\ref{d3cent} for the entropy of center  vortices.  Based on a simple picture\cite{corn070}  of the confining dynamics of a massless $q\bar{q}$  bound state, a good guess for the inverse  correlation length between bends of the pion world line, which is essentially the entropy per unit length,  is the inverse of the distance at which the confining energy is comparable to the quark mass, or $K_F/M(0)$.   This is quite comparable to the action per unit length, and should make an important {\em negative} contribution to the pion mass.   It is, in fact, indispensable if we ignore one-gluon effects, because the other contributions---the $K_Fr$ confinement energy and the kinetic energy---are positive.   

Another important entropic effect\cite{corn070,corn140} is the sprouting of long thin branches off the main pion trunk, representing the $\bar{q}q$ condensate that signals CSB.  Such a condensate can only form\cite{corn070} if the $q$ or $\bar{q}$ world lines have segments going backwards in time in Minkowski space, which is highly improbable for heavy quarks but not for massless ones.  The final picture of the pionic Wilson loop is that of a highly ramified and convoluted one with no large spatial separation $r$ between the $q$ and $\bar{q}$.  Such a Wilson loop is drastically different from the ones invoked earlier in Figs.~\ref{loop}, \ref{surfloop}, and seems to be\cite{antrib} in the equivalence class of branched polymers, with Haussdorf dimension 4.\cite{aikk}

It is not easy to deal directly with dynamical Wilson loops, highly-ramified or not (see an example\cite{shev} of such a CSB study, with further references).  Far better would be some form of Schwinger-Dyson equation that captures entropic effects more or less automatically.   Next in Sec.~\ref{areaent} we, as many others do,  use a simplified approximation to an area law expressed as an effective vectorlike propagator  $\sim 1/k^4$ coupled to the quarks.  When the quark worldline is closed, as it is in a Wilson loop, there is an  Abelian gauge invariance that has the remarkable effect of automatically regularizing the IR divergence of this propagator with a {\em physical } cutoff.  Moreover, it turns out that this physical regularization gives rises to a negative contribution to the pion energy that allows for CSB.

\subsection{\label{areaent}  An area law and entropy}

   In principle, the minimal-surface area law is expressible as a functional of the Wilson loop contour.    Only in  $d=2$ is this simple, because the minimal surface is flat:
\begin{equation}
\label{conflaw}
A_{\Gamma}=\oint\!\mathrm{d}z_i\,\oint\!dz_j'\,\frac{1}{4\pi}\delta_{ij}\ln |z-z'|.
\end{equation}
But in $d=3,4$ the contour form\cite{doug} of the minimal area is far from elementary, and difficult if not impossible to implement (see, however, studies\cite{corn131,makole}  of this problem).  So
we continue to use the simple form above, adjusting the overall factor as needed in $d=3,4$. 
In dimension $d$ the Fourier transform of the logarithmic propagator behaves like $k^{-d}$, and leads to IR singularities when used naively for such applications as CSB.   For example, the conventionally-defined static potential from this propagator can only be defined by supplying an infrared cutoff, and then this potential rises like $r$ as needed.   A closely-related issue is that Eq.~(\ref{conflaw}) has an Abelian gauge invariance that   has {\em nothing to do} with color gauge invariance.
If terms $\sim \partial_i\partial_j$ are added to the delta-function they give no contribution, because the contour is closed.      Among other things, this means that we need not specify the scale of $|z-z'|$ in the logarithm because changing it is a gauge transformation.  Without this gauge invariance there is no definite meaning to Eq.~(\ref{conflaw}).  Note that this potential, mimicking an area law, is always to be used to first order only in graphical applications.

The standard  gap equation for CSB refers to an open quark line: 
\begin{equation}
\label{naiveconf}
M(p^2)=\frac{1}{(2\pi )^4}\int\!\mathrm{d}^4k\,\gamma_{\mu}D^{eff}_{\mu\nu}\gamma_{\nu}(p-k)\frac{M(k^2)}
{k^2+M^2(k^2)}+\dots
 \end{equation}
where the dots stand for omitted gauge-dependent longitudinal terms.   
Postulating the $d=4$ version of Eq.~(\ref{conflaw}) for the effective propagator
\begin{equation}
\label{expreg}
D_{eff}(k)_{\mu\nu}\equiv \delta_{\mu\nu}D_{eff}(k);\quad D_{eff}(k)=\frac{8\pi K_F}{k^4}
\end{equation}
in this gap equation  violates the Abelian gauge invariance and requires an IR cutoff if CSB is to occur,  
  because if $M(0)\neq 0$ there is  an IR divergence in the integral, unless a special Abelian gauge is chosen.  There is also, of course, the problem that this gap equation refers to an amplitude that is not color-gauge-invariant either.  Whatever the Abelian or color gauge, one cannot be certain that the right physics is captured in those gauges.

 QCD physics should  be restricted to studies of {\em closed} world lines, corresponding to a physically-realizable $\bar{q}{q}$ pair such as describe the pion or the CSB condensate, and possessing both Abelian and color gauge invariance.    Nevertheless it should be   possible\cite{corn140} to extract an effective gap equation from a closed-loop\footnote{The gap equation is approximately what one would get by using the gauge-invariant pinch technique quark propagator\cite{cornbinpap}.} amplitude involving not only the massless quark undergoing CSB but also a hypothetical very heavy spectator quark.  Imposing these two gauge invariances has a perhaps unexpected bonus:  It automatically provides for a resolution of the IR divergence.   

First we take up the Abelian gauge invariance.   The sum of all closed-loop massive-particle amplitudes to a given order (in our case, first order)  in the propagator of Eq.~(\ref{expreg}) has this invariance. Call the sum of closed-loop graphs $F$, and construct the functional derivative
\begin{equation}
\label{abelgi}
G_{\alpha\beta}(k)=\frac{\delta F}{\delta D_{eff}(k)_{\alpha\beta}}.
\end{equation}
By Abelian gauge invariance this is conserved:  $k_{\alpha}G_{\alpha\beta}(k)=0$; a direct calculation verifies this.  A standard kinematic argument shows that if (as CSB demands) there are no massless lines in F, $G_{\alpha\beta}(k)$ must vanish at least quadratically at small $k$.  When $F$ is recovered by multiplying $G_{\alpha\beta}(k)$ by the effective propagator and integrating over $k$, all IR divergences are removed, and replaced by the lowest mass scale $M\equiv M(0)$  of the lines in $F$, where $M(0)$  is the CSB-generated running quark mass at zero momentum.  The would-be divergences are in effect ``regulated" by this mass.  A similar screening of an area law was found\cite{addav} in Coulomb gauge.  This sort of screening should not be confused with many applications\cite{saulbi} of a screened area law having nothing to do with closed loops or entropy, but simply used as a device to mimic string breaking.    An effective way\cite{corn140} of representing this physical regulation is to replace the original confining propagator by:
\begin{equation}
\label{newcong}
D_{eff}(k)_{\mu\nu}\equiv \delta_{\mu\nu}D_{eff}(k);\quad D_{eff}(k)=\frac{8\pi K_F}{(k^2+\tilde{m}^2)^2};
\end{equation}
we expect $\tilde{m}$ to scale with $M(0)$.  The static potential $V(r)$ for this propagator is now finite:
\begin{equation}
\label{newpot}
V(r)= -\frac{K_F}{\tilde{m}}e^{-\tilde{m}r} =  -\frac{K_F}{\tilde{m}}+K_Fr+\dots
\end{equation}
where the omitted terms should be irrelevant because large separations $r$ are never reached.  

The interesting thing about this equation is that the positive term $2M$ has been turned into a negative term!   We suggest\cite{corn140} that it represents entropy,  similar to way in which the entropy of polymers or center vortices is effectively a negative contribution to the mass, as shown in Eq.~(\ref{massdef}).The regulator term is negative, and we take this as a sign of entropic forces at work.  
To find  a useful estimate of $\tilde{m}$, consider the Hamiltonian $H=2M+V(r)+\dots $ for a color-singlet closed loop representing a $q\bar{q}$ bound state.      Long ago it was shown\cite{corn070}  that with the original confining propagator of Eq.~(\ref{expreg})  both the mass term $2M$   and the static potential $V(r)$ had IR divergences, their sum does not.   But only recently\cite{corn140} was the finite term remaining after cancellation of the divergences evaluted:
\begin{equation}
\label{mpot}
2M+V(r)=K_Fr-\frac{3K_F}{\pi M}.
\end{equation}
Once again, an unexpected negative term appears.  Equate the negative term in  Eq.~(\ref{mpot}) to that of the regulated static potential of Eq.~(\ref{newpot}) and find:
\begin{equation}
\label{regval}
\tilde{m}=\frac{\pi M}{3}\approx M.
\end{equation}
Now we have tamed the IR divergences of the effective confining propagator with the physical quantity $M$.

   These negative terms  are quite essential for CSB with only confining forces, since the pion must have zero mass.   A schematic Hamiltonian for the pion is:
 \begin{equation}
\label{hpi}
H_{\pi}\approx 2M+K_Fr+\frac{2}{r}\rightarrow \frac{2}{r}+K_Fr-\frac{3K_F}{\pi M}.
 \end{equation} 
Here we estimate the kinetic energies by $2/r$.  Without the negative (entropic) term a massless pion would be impossible, but with it there can be a massless bound state.  A simple minimization of $H_{\pi}$ on $r$ yields a zero eigenvalue at $M^2=9K_F/8$.

     Now we can construct\cite{corn140} a gap equation that is free of IR divergences,  simply by using for $D_{eff}(k)_{\mu\nu}$ the regulated propagator of Eq.~(\ref{newcong}), with $\tilde{m}\approx M$.  The solutions automatically break CSB, and yield $M^2\approx 0.8 K_F/\pi$,  or $M\approx$ 250 MeV (after the inclusion of about at 10\% increase from one-gluon exchange).  Note that attempting to go smoothly by variation of some parameter from CSB to chiral symmetry restoration is impossible; the only dimensionless parameter is the effective coupling $K_F/M^2$, which shows unbounded growth as one tries to take $M$ to zero.  This is quite unlike CSB with one-gluon exchange, where CSB is turned off by reducing the gluonic coupling.

These results for QCD have been verified\cite{natale1} and extended\cite{natale2} to a $[\bar{q}q]^2$ model for  CSB in technicolor models.

\section*{Acknowledgments}

I thank Martin Ammon for reading the manuscript.

\end{document}